\documentclass{amsart}

\makeatletter
\renewcommand{\email}[2][]{%
  \ifx\emails\@empty\relax\else{\g@addto@macro\emails{,\space}}\fi%
  \@ifnotempty{#1}{\g@addto@macro\emails{\textrm{(#1)}\space}}%
  \g@addto@macro\emails{#2}%
}
\makeatother

\usepackage{amsmath,amssymb,graphicx} 
\usepackage{amsaddr}
\usepackage{mathptmx} 
\usepackage{epstopdf}
\usepackage{natbib}

 \def\d{\textrm d} \def\ba#1\ea{\begin{align}#1\end{align}}

\DeclareMathAlphabet{\mathpzc}{OT1}{pzc}{m}{it}

\usepackage{mathptmx}
\DeclareMathAlphabet{\mathnew}{OMS}{cmsy}{m}{n}
\usepackage{mathtools} 
\def\mr{{R}}
\def\vth{\Theta}

\usepackage{mathrsfs}
\usepackage{graphicx}
\usepackage{amsmath}
\usepackage{epstopdf}
\usepackage{color}
\usepackage{soul}
\def\red{\textcolor{red}}
\def\blue{\textcolor{blue}}
\definecolor{darkblue}{RGB}{83,0,93}

\newsavebox{\astrutbox}
\sbox{\astrutbox}{\rule[-5pt]{0pt}{20pt}}

\newcommand\p{\ensuremath{\partial}}

\def\o{\over}
\def\p{\partial}
\def\be{\begin{eqnarray}}
\def\ee{\end{eqnarray}}
\def\bes{\begin{subeqnarray}}
\def\ees{\end{subeqnarray}}

\def\o{\omega}
\def\f{\frac}
\def\lp{\left(}
\def\rp{\right)}
\def\lb{\left[}
\def\rb{\right]}

\def\n{\nabla}

\def\ep{\epsilon}

\def\befi{\begin{figure}}
\def\eefi{\end{figure}}

\def\bce{\begin{center}}
\def\ece{\end{center}}

\def\d{\delta}

\def\th{\theta}

\def\L{\Lambda}

\def\d{\textrm{d}}
\def\n{\nabla}

\def\ba#1\ea{\begin{align}#1\end{align}}

\def\bsa#1\esa{\begin{subequations}
\begin{align}#1\end{align} \end{subequations}}

\usepackage{graphicx}
\usepackage{dcolumn}
\usepackage{bm}
\def\blue{\textcolor{blue}}                                                     
\def\red{\textcolor{red}}                                                       
                                                   
\def\black{\textcolor{black}}                                                   
                                                     
\definecolor{darkblue}{RGB}{83,0,93}                                            
\def\L{\mbox{------}}

\def\blue{\textcolor{blue}}                                                     
\def\red{\textcolor{red}}                                                       
                                                   
\def\black{\textcolor{black}}                                                   
  
\date{\today}

\numberwithin{equation}{section}
\usepackage[margin=1in]{geometry}

\begin{document}

\title[]{Broadband cloaking for flexural waves}


\author{Ahmad Zareei \ \ \& \ \ Mohammad-Reza Alam}
\address{Department of Mechanical Engineering, University of California, Berkeley}
\curraddr{}
\email{ahmad.zareei@gmail.com,reza.alam@berkeley.edu}
\thanks{}



\keywords{}

\date{}

\dedicatory{}

\begin{abstract}

The governing equation for elastic waves in flexural plates is not form invariant, and hence designing a cloak for such waves faces a major challenge.
Here, we present the design of a perfect broadband cloak for flexural waves through the use of a nonlinear transformation, and by matching term-by-term the original and transformed equations. 
 For a readily achievable flexural cloak in a physical setting, we further present an approximate adoption of our perfect cloak under more restrictive physical constraints.
Through direct simulation of the governing equations, we show that this cloak, as well, maintains a consistently high cloaking efficiency over a broad range of frequencies. 
The methodology developed here may be used for steering waves and designing cloaks in other physical systems with non form-invariant governing equations.

\end{abstract}

\maketitle

\section{Introduction}

The method of \textit{Transformation Optics}, originally developed in optics community for passive cloaking \cite[][]{Pendry2006,Leonhardt2006a}, offers a novel method for controlling electromagnetic waves using the subtle idea of coordinate transformation. Based on this method, invisibility cloaks for electromagnetic waves were designed, fabricated and successfully tested \cite[][]{Schurig2006c,Liu2009}. 

The most important necessary condition for applicability of the method  of transformation optics is that the governing equations must be \textit{form invariant} under coordinate transformation. Since physical systems admitting wave solutions share many common properties, it is well expected that the transformation optics method works in any wave system with form-invariant governing equation. This has been confirmed and cloaks have been designed and tested in a variety of other systems such as for acoustic waves \cite[][]{Cummer2007a, Chen2007a, Huang2014}, water waves \cite[][]{Chen2009,Berraquero2013, Zareei2015a} and matter waves \cite[][]{Zhang2008b}.

Flexural waves, such as those propagating on a thin elastic plates, have a governing equation that is known to be not form-invariant \cite[e.g.][]{Milton2006}. Therefore, the classical method of designing a cloak through transformation media method does not directly work in the context of flexural waves. One crude approximation is to adopt a form-invariant equation whose form is close to the governing equation of flexural waves and then use classical linear cloak design \cite[][]{Farhat2009c,Stenger2012}. 
While the resulting wave pattern about a to-be-cloaked cylinder may look like wave patterns of cloaking, a quantitative investigation of cloaking efficiency\footnote{Cloaking efficiency is defined as the ratio of \textit{reduction in the energy scattered to infinity in the presence of cloak} to \textit{energy scattered to infinity in the absence of the cloak}. A perfect cloak has an efficiency of 100\%.} \cite[][]{Norris1995a} 
%
%
reveals that such a cloak has a poor and in many cases even \textit{negative} cloaking efficiency (i.e. an object with the cloak about it scatters even more energy than the object without one, see also fig. \ref{fig5}, \ref{fig6}). 
Alternatively, if it is assumed that both density and elasticity of the material can be independently tuned, then a condition is obtained under which the highest order term of the governing equation satisfies the cloaking requirement \cite[][]{Brun2014}. This is theoretically an improvement, as the highest order term can be shown to play a more important role that the rest of the terms in the governing equation. Nevertheless, fabricating a material with a variable density \textit{and} elasticity is a serious challenge \cite[][]{Stenger2012}. Along the same line, more degrees of freedom such as several independent elastic parameters may be assumed to improve the theoretical performance \cite[][]{Colquitt2014a}, but this makes the realization of the cloak in physical space even farther from achievable. 



Here we present the design of a perfect broadband cloak for flexural waves. For the cloak to be realizable in the physical domain, we put the constraint that the density $\rho$ is constant and only the modulus of elasticity $E$ can be changed across the cloak. We employ a nonlinear transformation and, by choosing proper material properties and pre-stressing, match term-by-term the coefficients in the original and transformed equations. {We show rigorously that the transformed equation matches perfectly with the orthotropic and inhomogeneous plate's equation.} 

\section{Governing Equations}

For an isotropic plate with thickness $h$ and density $\rho_0$, governing equation for out of the plane displacement 
$\eta(\mr,\vth,t)$ 
in the $z$ direction normal to the plate's surface reads \cite[e.g.][]{Timoshenko1940}
\begin{align} \label{eq1}
   {D_0}\Delta^2 \eta + \rho_0 h \eta_{tt}=0,
\end{align}
where $D_0 = E_0 h^3/12(1-\nu^2)$ is the flexural rigidity, $E_0$ is the Young Modules, $\nu$ is the Poisson ratio, and $\Delta$ is the horizontal Laplacian operator  {in $(\mr,\vth)$ directions}.

To cloak a circular region $A_c$ (radius $a$) with a cloak of outer radius $b$ co-centered with $A_c$, we need to map the  {region of $0\leq \mr\leq b$ to the cloaking region $a\leq r\leq b$}. We use the nonlinear transformation $ \mathcal{F}$ defined as
\begin{eqnarray}  \label{eq2}
  \mathcal {F} : \left\{
    \begin{array}{ll}
      r= \sqrt{\lp 1- {a^2}/{b^2}\rp \mr^2 + a^2}, & 0\leq \mr \leq b, \\
      \th = \vth,
    \end{array}
\right.
\end{eqnarray}
that has a special property of its Jacobian being a constant \cite[][]{Zareei2015a}. Using this transformation and further assuming a time-periodic motion of frequency $\o$, equation \eqref{eq1} is maped to \cite[using Lemma 2.1 in][]{Norris2008}
\begin{align}\label{eq3}
  D_0\tilde\nabla^2 \tilde \nabla^2 \eta - {\rho_0 h \omega^2} \eta = 0,
\end{align}
where
\begin{align}\label{eq4}
  \tilde\nabla^2 =   \lp 1-\f{a^2}{b^2} \rp\lb \f{1}{r} \f{\p}{\p r}\lp \f{r^2-a^2}{r} \f{\p}{\p r}\rp + \f{1}{r^2-a^2} \f{\p^2}{\p \th^2} \rb.
\end{align}
Note that if $a=0$, then $\tilde\nabla^2\equiv \Delta$. 

In a traditional cloak design for form-invariant governing equations \cite[e.g.][]{Pendry2006}, material properties as functions of spatial variables are determined such that the transformed equation \eqref{eq3} with the new material properties becomes equivalent of the original equation \eqref{eq1}. If we do the same here, {the rigidity $D$ {becomes spatially variable in different directions}, which means the required material for cloaking is \textit{inhomogeneous} and \textit{orthotropic}.} The issue is, equation \eqref{eq1} with $D=D(r,\theta)$, is \textit{not} the governing equation for an inhomogeneous and orthotropic plate. In fact, the governing equation for a general $D(r,\theta)$ is very much different in the look [equation \eqref{923} in the Appendix], and most importantly this equation is \textit{not} form-invariant.



With this knowledge, we therefore look for material properties that result in the matching of the coefficients  of the two equations (i.e. \eqref{eq3} and \eqref{923}). We find that if we choose the following material parameters 
\begin{subequations} \label{eq5}
\begin{align}
  &D_r = \alpha^2 \mathcal{A}^2(r)  D_0, \label{eq5a}\\
  &D_\th = \alpha^2 \lp1/\mathcal{A}(r)\rp^2 D_0, \label{eq5b}\\
  &D_{r\th}= \alpha^2 D_0, \\
  & \nu_\th = \f{1}{\alpha^2 \mathcal{A}^2(r)}\lb \mathcal{B}(r) - 4 \log\mathcal{A}(r) \rb,
\end{align}
\end{subequations}
where $\mathcal{A}(r) = 1-\lp a/r\rp^2 $, $\alpha= 1-\lp a/b\rp^2 $ and $\mathcal{B}(r)  = 3({r}/{a})\log\lb {(r-a)}/{(r+a)}\rb - 2{a^2}/{(r^2-a^2)}$  then between equation \eqref{eq3} and \eqref{923} all terms that include 4th order derivatives (i.e. highest order appearing in these equation), all 3rd order, 2nd order and 1st order terms {match perfectly} except two extra terms in the transformed equation which are factors of $\p^2\eta / \p r^2$ and $\p \eta/\p r$. Intrestingly, these extra terms go to zero if the penetration depth is small (see Eq. \eqref{930} in the Appendix). Alternatively, these two extra terms can be handled with a material that is pre-stressed in the radial direction (i.e. $N_{\th\th} = N_{r\theta} =0$) with a radial body force (i.e. $S_\th=0$) according to
\begin{subequations}
  \begin{align}
    & \f{N_{rr}}{D_0} = \f{1}{2a^2}\lb  \lp \f{a}{r}\rp^8  \mathcal{N}(r) + \f{3}{2}\lp\f{a}{r}\rp\log \lp\f{r-a}{r+a} \rp\rb  ,\\
    & \f{S_r}{D_0}  = -\f{3}{a^3}\lp \f{a}{r}\rp^{11}\mathcal{S}(r)
  \end{align}
\end{subequations}
with $\mathcal{S}(r) = \lb 5-12(r/a)^2+8(r/a)^4\rb/\lb 1-(a/r)^2 \rb^2$ and $\mathcal{N}(r) = \lb 6-10(a/r)^2-2(r/a)^4 + 3(r/a)^6 \rb/ \lb 1-(a/r)^2\rb$. 

The above cloak for flexural waves is a rigorously derived \textit{perfect} (i.e. efficiency is theoretically unity) and \textit{broadband} cloak. We now move to the next level by designing an approximate adoption of this perfect cloak restricted by more physical constraints that make achieving such a cloak even easier in an experimental setting. Specifically, the goal is to find an approximate adoption of our cloak that only requires \textit{concentric layers of homogeneous material} \cite[c.f.][]{Stenger2012, Farhat2009c}. Using these concenteric isotropic layers, we can only achieve a variable radial and azimuthal flexural rigidities. 

\begin{figure}
 \centering
 \hspace{-2mm}\includegraphics[width=3.5in]{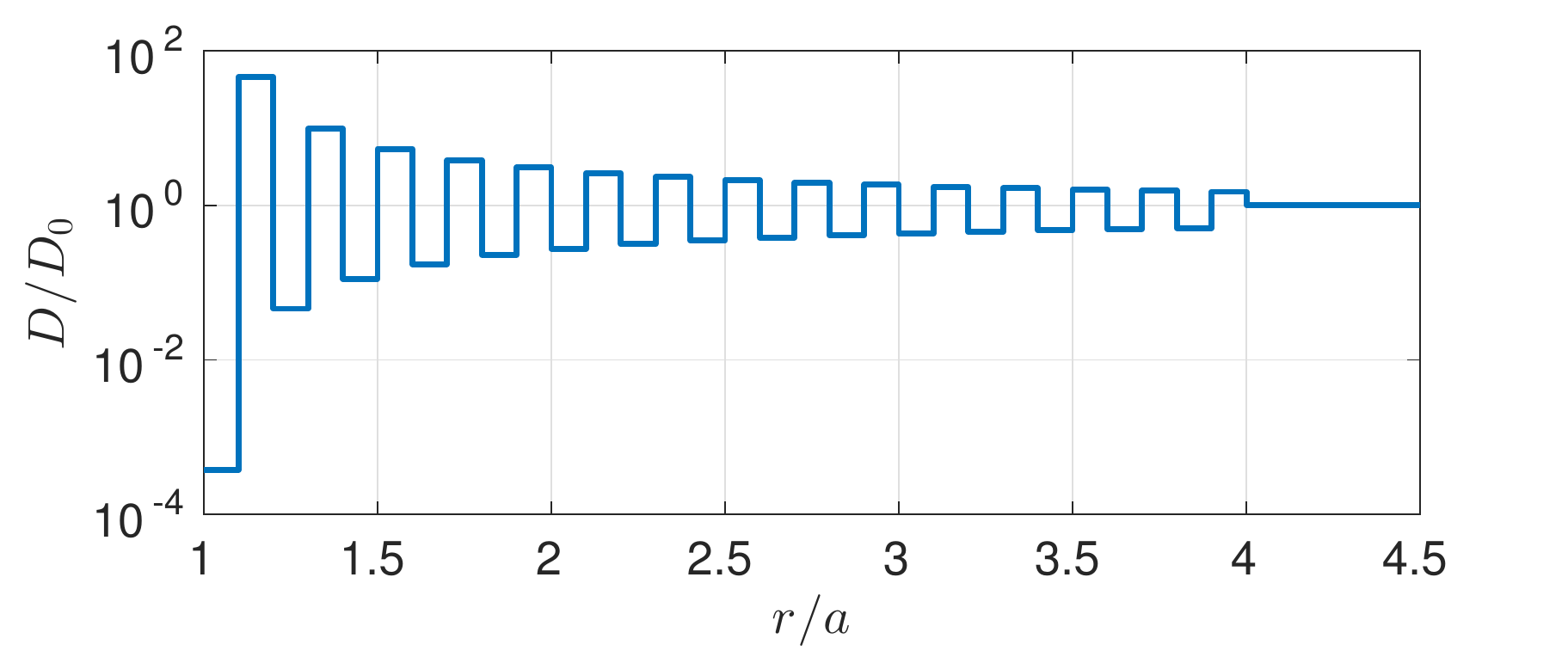}
 \caption{Profile of the rigidity as a function of $r$ required to achieve a cloak for flexural waves. Each layer is made up of a homogeneous and isotropic material, but the averaged properties provides an inhomogeneous and orthotropic apparent rigidity according to \eqref{eq5}. In the design presented here, the number of layers is $15$, each layer is divided to two sub-layers, and $b/a=4$. 
%
 }\label{fig1}
\end{figure}

Assuming only a variable radial and azimuthal rigidities in our cloak (i.e. equations \eqref{eq5a} and \eqref{eq5b}), only the highest derivatives (i.e. 4th order and 3rd order terms) match between equations \eqref{eq3} and \eqref{923}. In order to test the effectiveness of our cloak which is based on the nonlinear transformation \eqref{eq2}, in comparison with the one based on linear transformations \cite[][]{Stenger2012, Farhat2009c}, we use $N=15$ layers of homogeneous but orthotropic materials, that is, $D_r,D_\theta$ are  {constant throughout each layer}. To achieve the required orthotropic response, each layer is divided into two sub-layers of isotropic and homogeneous materials with different  {rigidities}. These two sub-layers can be shown through homogenization that provide the required orthotropic properties \cite[e.g.][]{Cheng2008b}. In implementing the cloak, since according to \eqref{eq5} the rigidities go unbounded at the inner boundary, a small offset is introduced such that the rigidity of the first layer (next to $r=a$) is calculated at this offset distance  from the inner boundary. This offset can be shown to be equivalent of transforming a region of $\varepsilon \leq R \leq b$ to the cloaking region $a\leq r \leq b$ {[see Eq. \eqref{995} and \eqref{996} in the Appendix]}. In the numerical simulations that follow, we choose this offset to be $15\%$ of thickness of a layer.  The final profile of the rigidity of the isotropic and homogeneous layers for the case of $b/a=4$ is shown in fig. \ref{fig1}.

\section{Numerical Simulation \& Results}

In order to numerically solve the thin plate's equation with flexural rigidity as shown in fig. \ref{fig1}, we use the spectral method with Fourier expansion in the azimuthal direction and Bessel functions in the radial direction. Assuming time harmonicity of $\omega$, the solution in each layer is therefore expressed as $\text{Re} \lb \eta (r,\th) \exp (i \omega t)\rb$. We expand the spatial part $\eta(r,\th)$ as
\begin{align}
     \eta (r,\th) =  \sum_{n=-\infty}^{\infty} \eta_n(r)  \exp \lp {in\th}\rp,
\end{align}
where $\eta_n(r) = A_n J_n (k_i {r}) + B_n I_n(k_i {r}) + C_n Y_n(k_i {r}) + E_nK_n (k_i r)$ with $k_i^4 = \rho_0 h \omega^2/D_i $ and $D_i$ being the flexural rigidity of the layer. Here, $J_n(.), Y_n(.) $ and $I_n(.), K_n(.) $ are respectively Bessel and modified Bessel functions of the first and second kind. Also  $A_n, B_n, C_n$ and $E_n$ are coefficients that are later found by satisfying the boundary conditions. These boundary conditions are continuity of displacement $\eta$, its radial derivative $\eta_r$, momentum $M$ and shear force $V$ at the boundaries {[see Eq. \eqref{4417} and \eqref{4418} in the Appendix]}. Note that spatial part of the incident planar wave can be written as $\eta^{inc} = a_0 \exp \lp i k_0 x \rp =  a_0 \sum_{n=-\infty}^{\infty} i^n J_n(k_0 r)\exp(i n \th)$, where $a_0$ is the amplitude of the wave, $k_0^4 = \rho_0 h \omega^2/D_0$ and $D_0$ is the constant flexural rigidity outside the cloak.

In order to quantitatively analyze the efficiency of the cloak, we calculate the scattering cross section which corresponds to the energy scattered to the infinity. The scattering displacement field is $\eta^{sc} = \eta - \eta^{inc}$, where $\eta^{inc}$ is the incident plane wave. The scattered far field amplitude $f(\th)$ is defined through \cite[see e.g.][]{Norris1995a}
\begin{align}
  \eta^{sc} = \f{a_0}{\sqrt{2r}} e^{i \lp {k_0 r - \pi/4}\rp } f(\th) + \mathcal{O}(1/\sqrt{r})
\end{align}
and the total scattering cross section is $\sigma^{sc} = {1}/{2\pi} \oint |f(\th)|^2\d \th$. 

\begin{figure}
\centering
\begin{tabular}{c}
\hspace{-5mm}\vspace{-0mm}\includegraphics[width=3.75in]{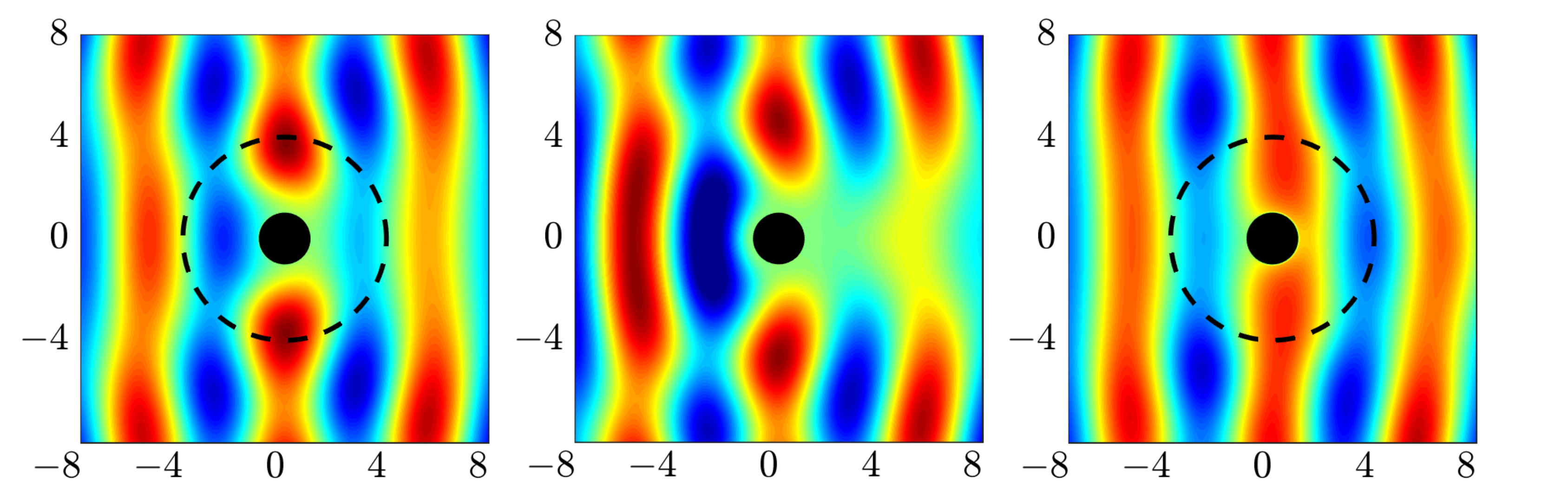}\\
\hspace{-5mm}\vspace{-0mm}\includegraphics[width=3.75in]{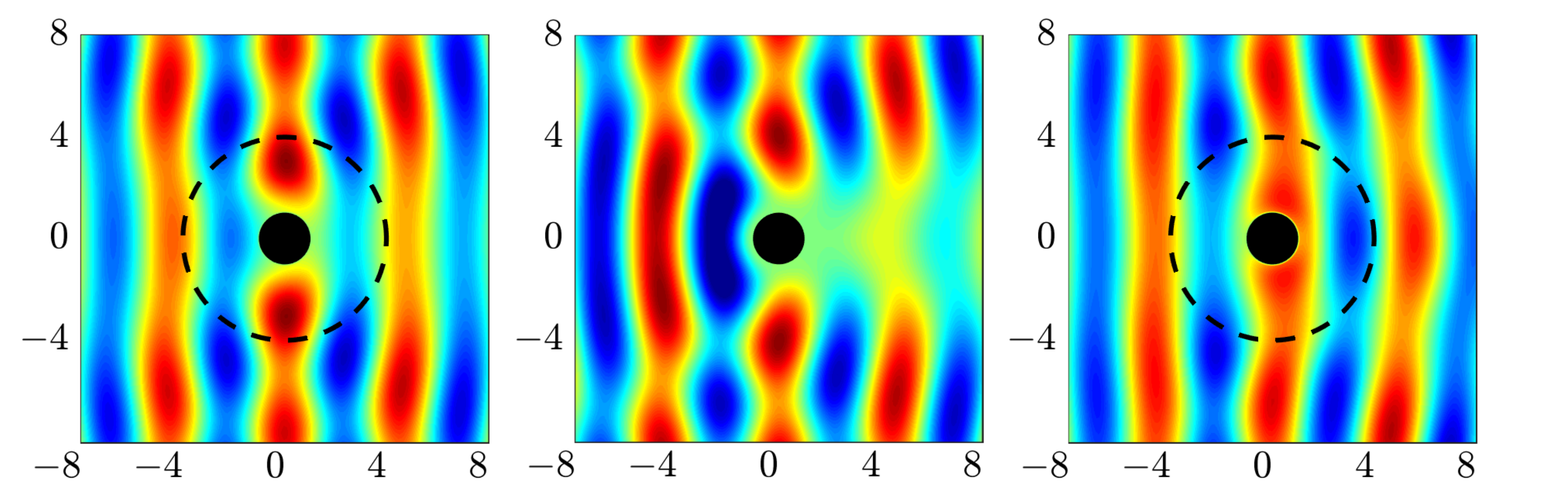}\\
\hspace{-5mm}\vspace{-0mm}\includegraphics[width=3.75in]{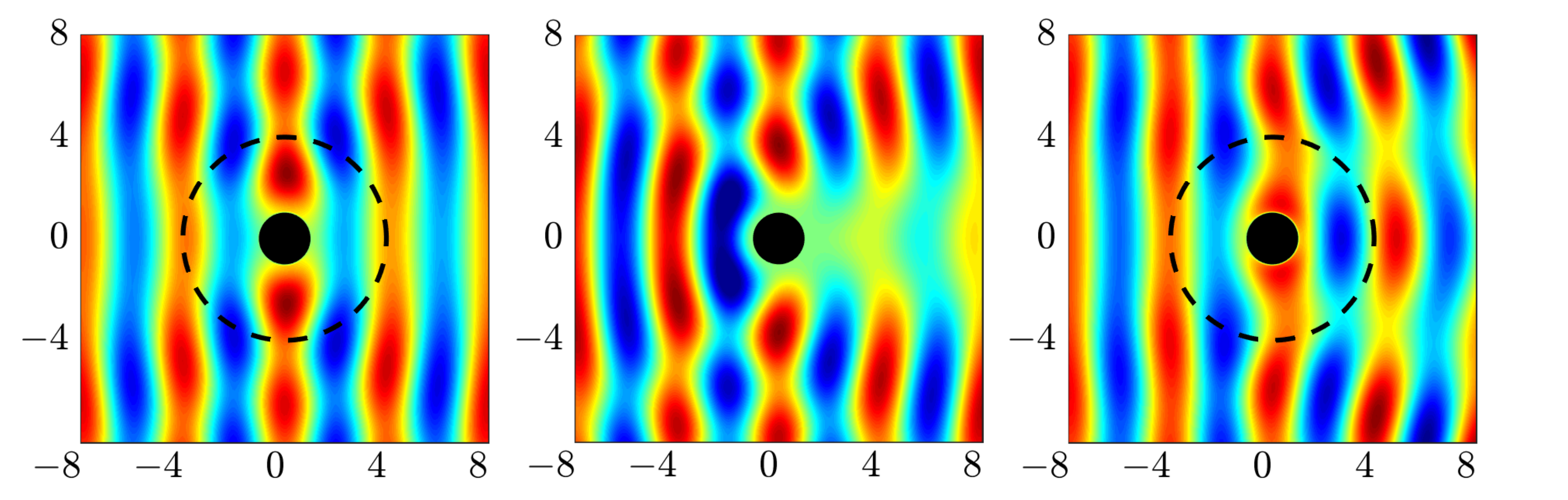}\\
\hspace{-5mm}\vspace{-0mm} \includegraphics[width=3.75in]{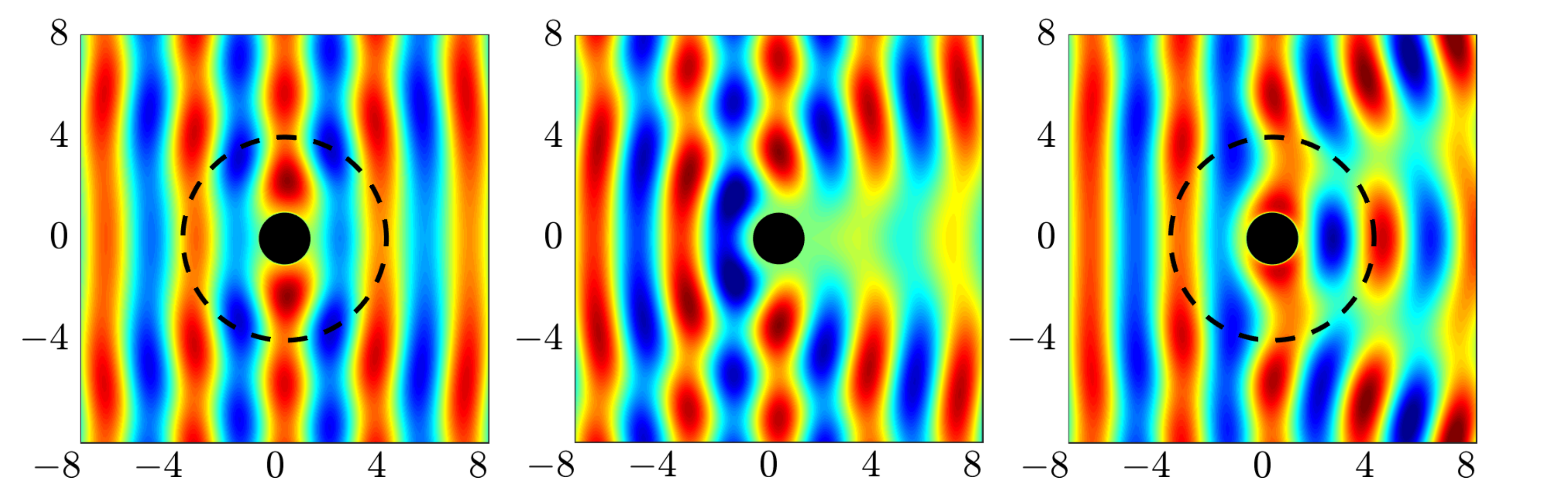}\\
\hspace{-5mm}\vspace{-0mm} \includegraphics[width=2.75in]{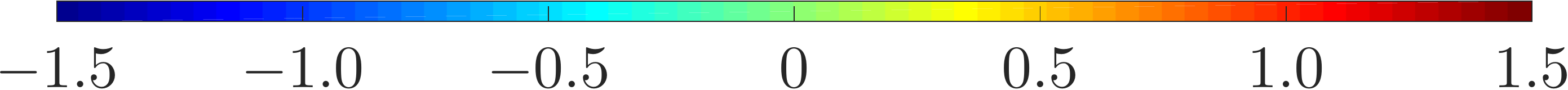}
\end{tabular}
 \caption{Nondimensionalized displacement field $\eta/a_0$ for a nonlinear cloak (left column), without any cloak (middle column), and linear cloak (right column). Each row corresponsds to a different frequencies: $200$Hz (first row), $300$Hz (second row), $400$Hz (third row) and $500$Hz (the last row). Coordinates are nondimensionalized with radius of the inner cylinder $a$ and the cloak size for both linear and nonlinear cloak is $b/a=4$. The cloaks are approximated with $N=15$ layers of homogeneous anisotropic materials with each layer composed of two sub-layers made up of different homogeneous and isotropic materials. For a direct comparison with Stenger et al. \cite{Stenger2012}, Frequencies are obtained using the values of $h=1mm$, $a=1.5cm$, $\rho = 2000kg/m^3$ and $D_0=0.1037 Nm^2$. A nonlinear cloak shows a consistent cloaking efficiency for different frequencies, while performance of the linear cloak drops significantly as the frequency increases. For a quantitative comparison of performance, see figures \ref{fig5},\ref{fig6}.}\label{fig4}
\end{figure}

We present here a side-by-side comparison of surface elevation $\eta$ and scattered far field amplitude $f(\theta)$ for three cases: i. in the absence of cloak, ii. with the claimed linear cloak of \cite[][]{Farhat2009c, Stenger2012}, and iii. with our nonlinear cloak. We implement the linear cloak according to Eq. (4) of \cite{Farhat2009c} for a cloak size of $b/a=4$. We approximate the cloak $N=15$ concentric layers that are homogeneous but anisotropic and then we use two isotropic and homogeneous sub-layers to approximate each of the 15 layer \cite[see e.g.][]{Farhat2009c, Stenger2012, Cheng2008}. The resulted layers of isotropic and homogeneous materials approximates the anistropic inhomogeneous cloak. We do the same for the nonlinear cloak but in this case according to equation \ref{eq5}. The result for a linear cloak design \cite[][]{Farhat2009c, Stenger2012} and the nonlinear cloak for the range of the frequencies of $f=200$Hz$-500$Hz alongside with the case when there is no cloak is shown in fig. \ref{fig4} for comparison. 

Looking at the displacement field in fig \ref{fig4}, both of the cloaks seems to be effective at lower frequencies. In fact, the linear cloak \textit{looks} to be more effective than the nonlinear cloak in $f$=200Hz. More specifically, the nonlinear cloak appears to have less scattering downstream of the cylinder compared to the linear cloak and the linear cloak have less scattering upstream of the cylinder compared to the nonlinear case. By increasing the frequency, we observe that downstream scattering of the nonlinear cloak is much better in preserving the wave shape; while at the upstream of the cylinder the linear cloak has less scattering. 

\begin{figure}
\centering
\begin{tabular}{c}
\hspace{-4mm}\includegraphics[width=1.65in]{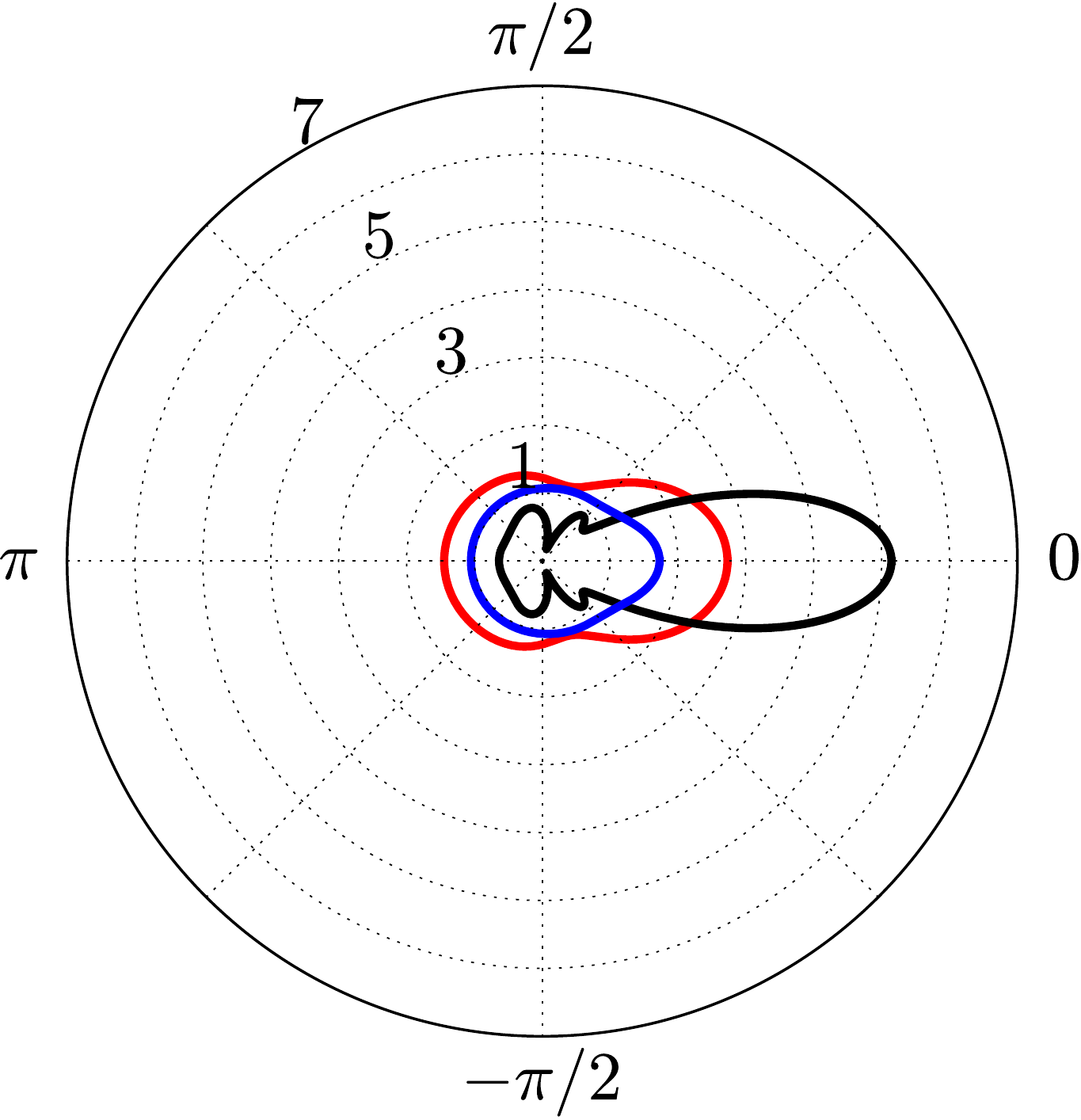}
\includegraphics[width=1.65in]{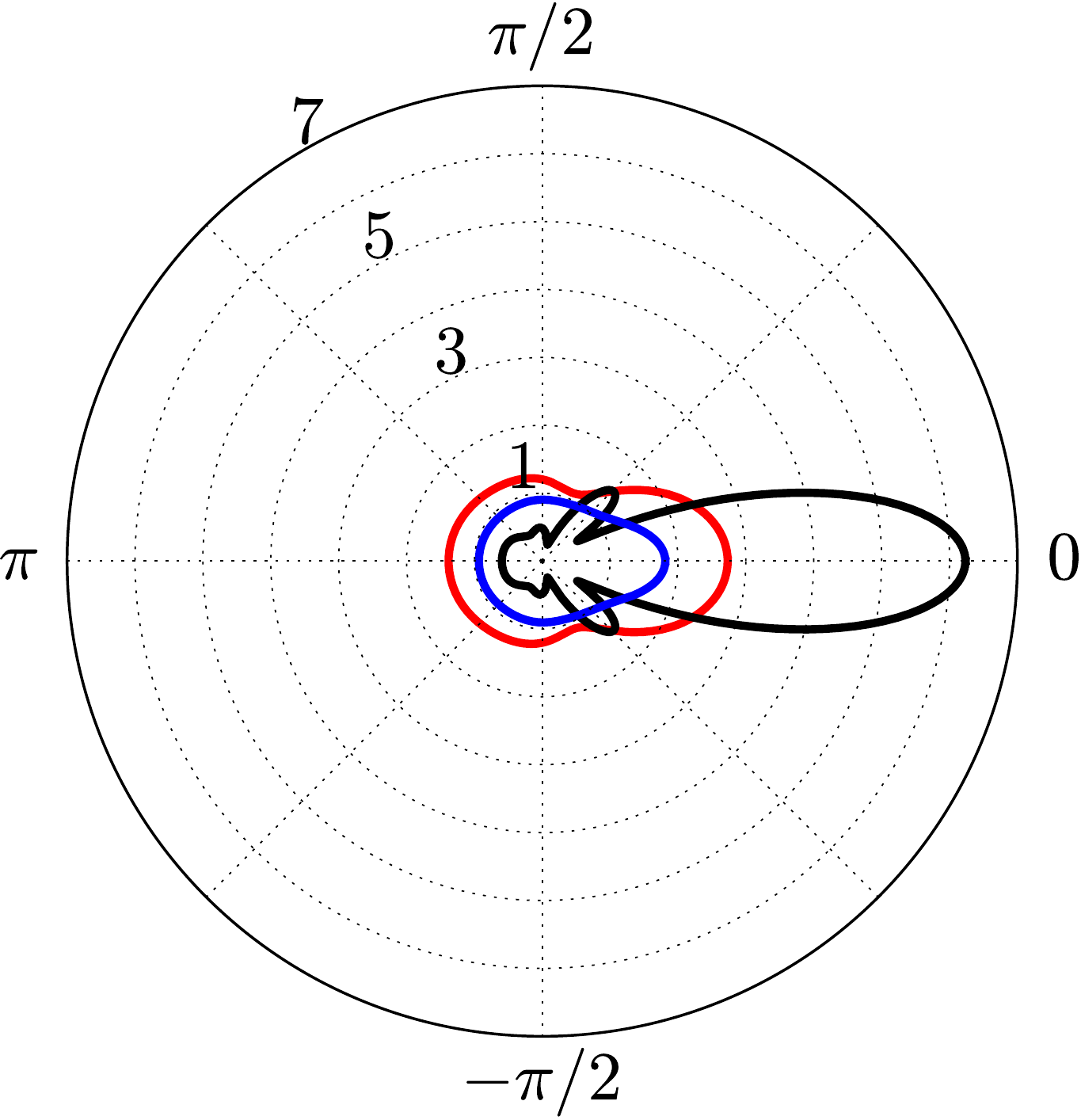}
\put(-240,110){$(a)$}
\put(-120,110){$(b)$}
\\
\hspace{-4mm}\includegraphics[width=1.65in]{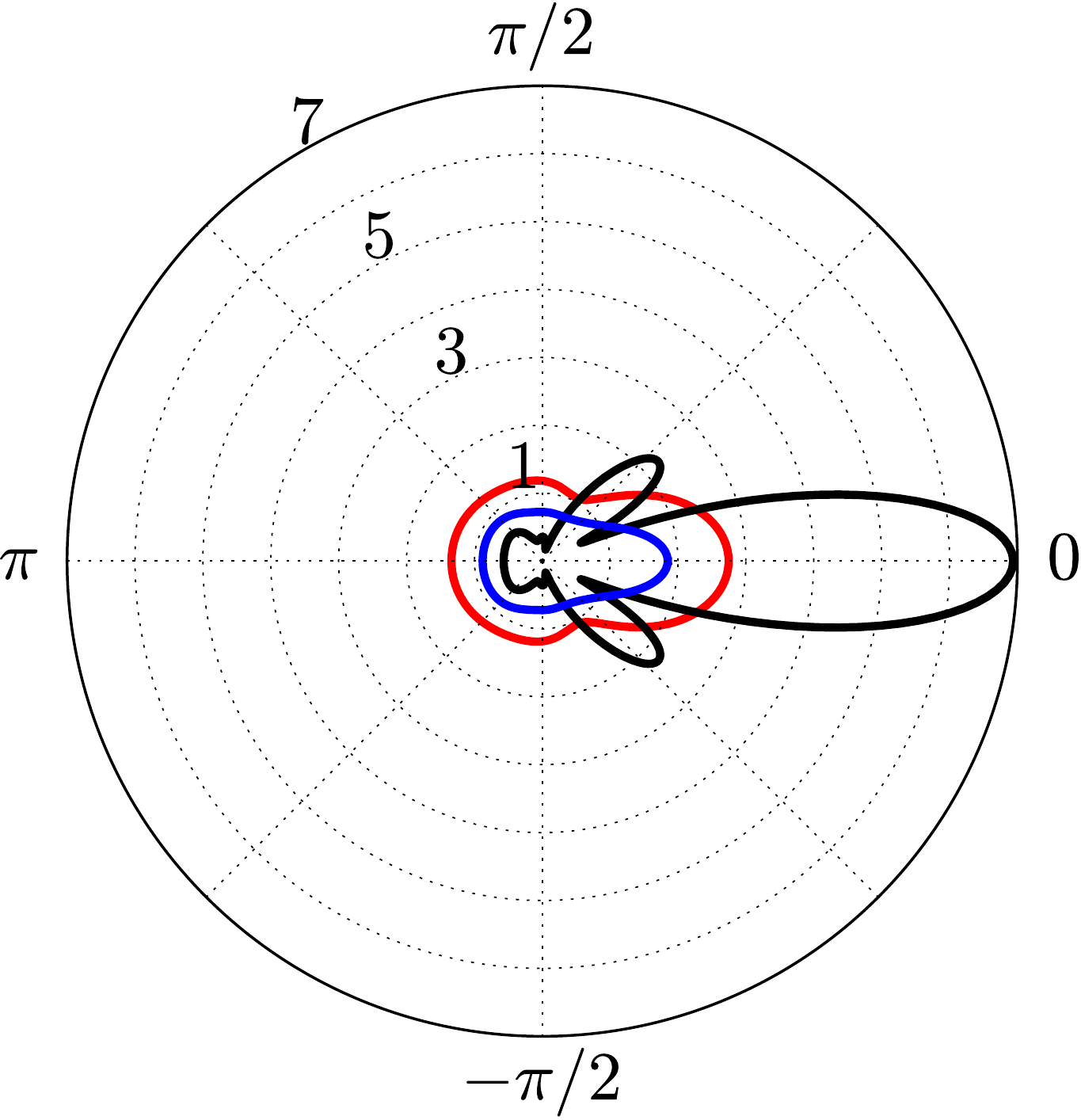}
\includegraphics[width=1.65in]{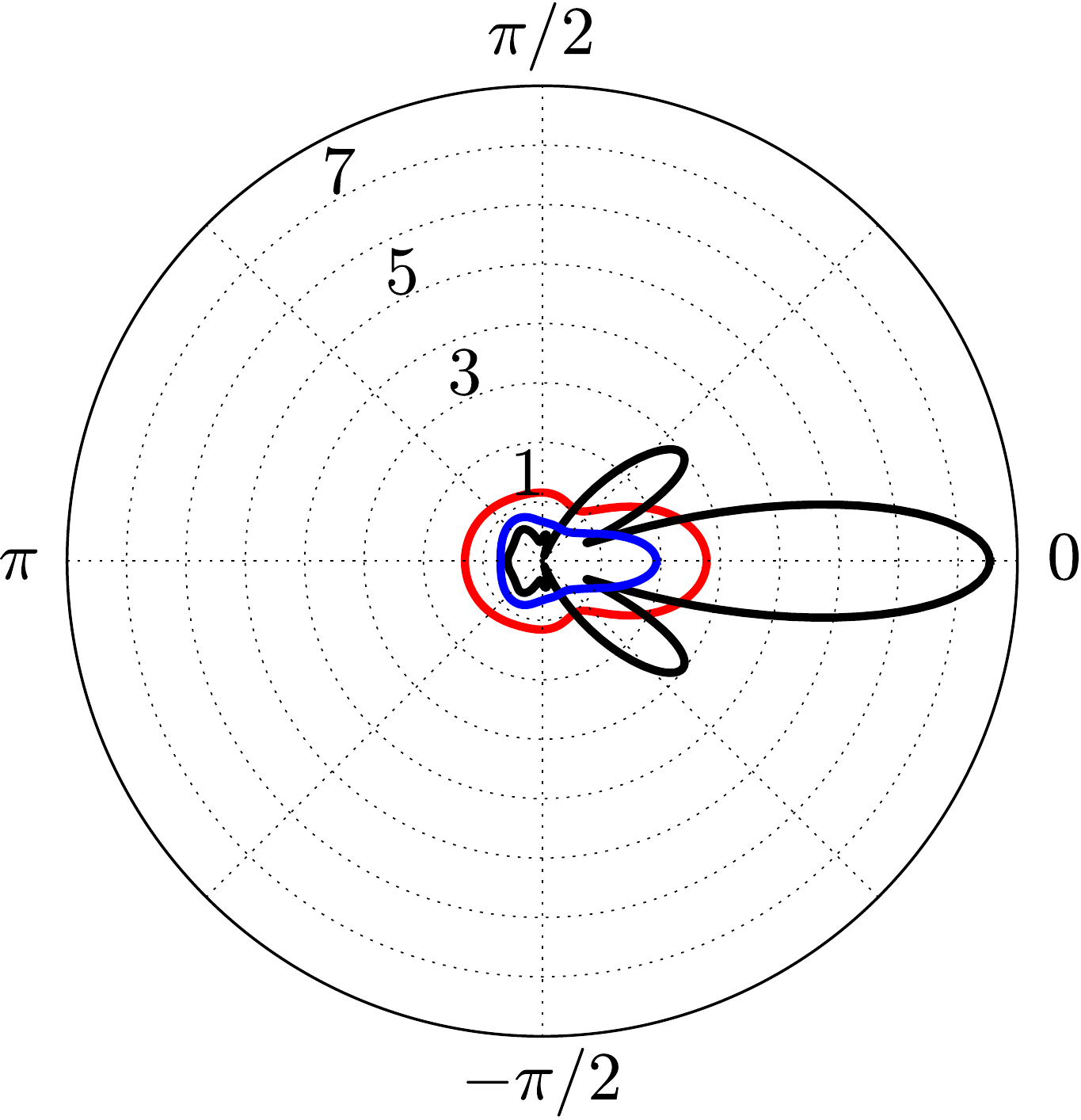}
\put(-240,110){$(c)$}
\put(-120,110){$(d)$}
\end{tabular}
 \caption{Polar plot of absolute value of scattering amplitude $|f(\th)|$ for the linear cloak (\black{\L}), without cloak (\red{\L}), and with nonlinear cloak (\blue{\L}) for different frequencies of (a) $200$Hz, (b) $300$Hz, (c) $400$Hz and (d) $500$Hz.}\label{fig5}
\end{figure}

In order to quantitatively test the effectiveness of the cloaks, we present in fig. \ref{fig5} the absolute value of the scattering amplitude $|f(\th)|$ at different angles for both of the cloaks at different frequencies $f=200$Hz$-500$Hz. As is seen, in all of the frequencies, although in upstream of the cylinder the linear cloak is reducing the amount of energy scattered to infinity compared to the nonlinear case, far more energy is scattered in the downstream of the cylinder in the linear cloak, even larger than the case when there is no cloak. Therefore, in all cases the linear cloak scatters \textit{more} energy to the downstream compared to when the cloak does not exist. Our nonlinear cloak consistently achieves a lower scattering in all angles with no exception.

\begin{figure}
\centering
\begin{tabular}{c}
\hspace{-4mm}\includegraphics[width=3.5in]{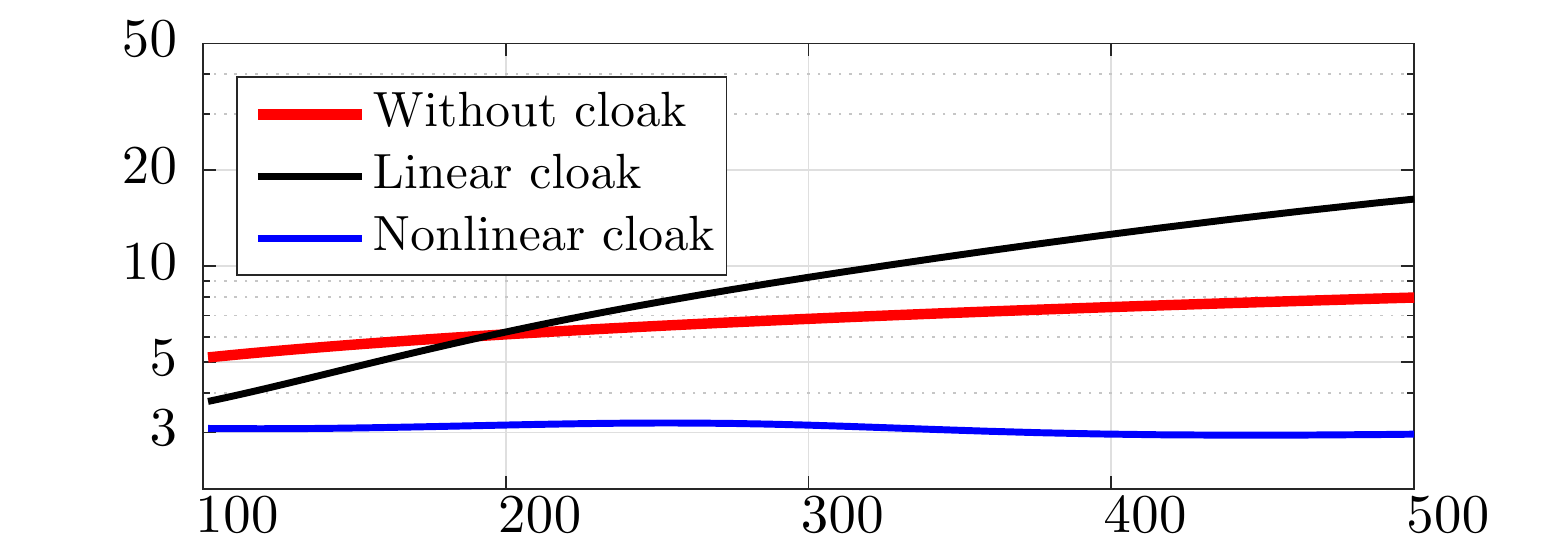}
\put(-135,-4){$f$}
\put(-243,40){\rotatebox{90}{$k\sigma^{sc}$}}
\end{tabular}
 \caption{{Nondimensionalized total scattering cross section with wavelength i.e. $k\sigma^{sc}$} for the a range of frequencies for 3 different cases of without cloak, with a linear cloak and with a nonlinear cloak.}\label{fig6}
\end{figure}

In order to quantitatively see the net effect of the cloak in terms of the total energy scattered to the infinity, we also calculate and plot the total scattering cross section $\sigma^{sc}$ (fig. \ref{fig6}). In the frequencies bellow $200$Hz, although the linear cloak scatters more energy at certain angles (at the downstream side of the cylinder) compared to the case with no cloak [see fig \ref{fig5}], the total energy scattered to the infinity is smaller than the case with no cloak. At frequencies above $200$Hz the linear cloak both scatters far more energy at the back of the cylinder and also in total. Note that, the total scattering cross section for the nonlinear cloak stays always smaller than the case without any cloak. This underlines the broadband effectiveness of our proposed nonlinear cloak. 
%


In summary, we presented here the design of a perfect broadband cloak for flexural waves. Since the governing equations for flexural waves are not form-invariant, the traditional cloak design methodology through linear transformation optics scheme does not apply here. We therefore employed a nonlinear transformation and matched, term-by-term, the transformed equation with the true governing equation for an inhomogeneous and orthotropic plate equation. We {showed rigorously that the resulting cloak is perfect. We also presented an adoption of our perfect cloak obtained under more restrictive physical constraints that make the design more amenable for experimental investigations. These constraints are that cloak can only include a finite number of concentric layers of homogeneous materials and that only modulus of elasticity can be variable from layer to layer. We presented this approximate cloak, and showed via direct simulation that this experimentally realizable cloak of such type has a consistent performance in all spatial directions, and also has a broad bandwidth of high efficiency.}


The nonlinear cloak proposed in here is combination of layers of homogeneous and isotropic materials, which are amenable to physical fabrication and testing and real-life application \cite[e.g. potentially in cloaking against earthquakes][]{Brule2014, Colombi2015b}.
The nonlinear transformation proposed here, may be applied for other types of the waves to soften the required material properties. For instance, in electromagnetism, with this nonlinear cloak, we can remove one degree of the freedom and keep permeability (permitivity) as a constant in cloaking for transverse magnetic (electric) waves. 

\section{Appendix}

\section*{Governing Equation}

Assuming an orthotropic and inhomogeneous plate, under pure bending and in the absence of in-plane forces, we have \cite[][]{Aeronautics}
\begin{align}\label{911}
  \f{\p^2 M_R}{\p R^2} + \f{2}{R} \f{\p M_R}{\p R}+ \f{2}{R} \f{\p^2 M_{\Theta R}}{\p R \p \Theta} + \f{2}{R^2} \f{\p M_{\Theta R}}{\p \Theta} + \f{1}{R^2} \f{\p^2 M_\Theta}{\p \Theta^2} -\f{1}{R} \f{\p M_\Theta}{\p R} + \rho_0 h \omega^2 \eta = 0,
\end{align}
where $M_R, M_\Theta$ and $M_{R\Theta}$ are the bending moments, $\rho_0$ is the density of the plate, $h$ is the thickness and $\omega$ is the frequency of the wave. Bending moments $M_R, M_\Theta$ and $M_{R\Theta}$ are found as
\begin{subequations}
\begin{align}
  M_R &= -D_R \lb \f{\p^2 \eta}{\p R^2} + \nu_\Theta \lp \f{1}{R} \f{\p \eta}{\p R} + \f{1}{R^2} \f{\p^2 \eta}{\p \Theta^2} \rp\rb, \\
  M_\Theta &= -D_\Theta \lp \f{1}{R} \f{\p \eta}{\p R} + \f{1}{R^2} \f{\p^2 \eta}{\p \Theta^2}  + \nu_R \f{\p^2 \eta}{\p R^2} \rp, \\
  M_{R\Theta} &= -2D_K \f{\p}{\p R} \lp\f{1}{R} \f{\p \eta}{\p \Theta} \rp,
\end{align}
\end{subequations}
where $\eta({R,\Theta})$ is the out-of-plane displacement,  $D_R, D_\Theta$ are the flexural rigidities in the $R,\Theta$ directions respectively, $D_K$ is the shearing rigidity and $\nu_R, \nu_\Theta$ are the Poisson ratios in the radial and tangential directions. Note that the radial and tangential rigidities $D_R, D_\Theta$ satisfy the symmetry relation $D_R \nu_\Theta = D_\Theta \nu_R$. Using these relations and defining $D_{R \Theta} = 2D_K + D_R \nu_\Theta$, equation \eqref{911} simplifies to \cite[][]{Brun2014}
\begin{align}\label{923}
  &D_R \f{\p^4 \eta}{\p R^4}  + \f{D_\Theta}{R^4} \f{\p^4 \eta}{\p \Theta ^4} + \f{2 D_{R\Theta}}{R^2}\f{\p^4 \eta}{\p R^2 \p \Theta^2} +\lp \f{2D_R}{R} + 2\f{\p D_R}{\p R} \rp \f{\p^3 \eta}{\p R^3} \nonumber \\
\quad & + \lp \f{2}{R^2}\f{\p D_{R\Theta}}{\p R}-\f{2 D_{R\Theta}}{R^3} \rp \f{\p ^3 \eta}{\p R \p \Theta^2} + \lb \f{1}{R^2}\f{\p}{\p R}\lp R^2 \f{\p D_R}{\p R}\rp + \f{1}{R} \f{\p (D_R \nu_\Theta)}{\p R}-\f{D_\Theta}{R^2}  \rb \f{\p^2 \eta }{\p R^2} \nonumber \\
\quad & +\lp \f{2D_{R\Theta}}{R^4}  -\f{2}{R^3} \f{\p D_{R\Theta}}{\p R}+ \f{2D_\Theta}{R^4} -\f{1}{R^3} \f{\p D_\Theta}{\p R} +\f{1}{R^2}\f{\p^2 (D_R \nu_\Theta)}{\p R^2}\rp \f{\p^2\eta}{\p \Theta^2} \nonumber \\
\quad & + \lp \f{D_\Theta}{R^3} -\f{1}{R^2}\f{\p D_\Theta}{\p R} +\f{1}{R} \f{\p^2 (D_R \nu_\Theta)}{\p R^2} \rp \f{\p \eta}{\p R} - \rho_0 h \omega^2 \eta =0.
\end{align}
Assuming a constant rigidity $D_R  = D_\Theta = D_{R\Theta} = D_0$, equation \eqref{923} simplifies to the famous biharmonic plate's equation as
\begin{align}\label{924}
  D_0\Delta^2 \eta - \rho_0  h \omega^2 \eta = 0.
\end{align}
Now as we aim to cloak a circular region $A_c$ with radius $a$, with a cloak of outer radius $b$ co-centered with $A_c$, we use the following transformation $\mathcal{F}$  to map the area $0 \leq R\leq b$ to the cloaking region $a\leq r\leq b$ \cite[][]{Zareei2015a}
\begin{eqnarray}  \label{925}
  \mathcal{F}: \left\{
    \begin{array}{ll}
      r=f(R) = \sqrt{\lp 1- {a^2}/{b^2}\rp R^2 + a^2}, & 0\leq R \leq b, \\
      \th = \Theta.
    \end{array}
\right.
\end{eqnarray}
The Jacobi of the transformation $\mathcal{F}$ in the polar coordinate is
\begin{align}
  \mathbf{F} =     \sqrt{1-a^2/b^2}  \lp
  \begin{array}{cc}
    {\sqrt{r^2-a^2}}/{r} & 0 \\
    0 & {r}/{\sqrt{r^2-a^2} }
  \end{array}
\rp_{(\mathbf{e}_r, \mathbf{e}_\th)},
\end{align}
where $\left\{\mathbf{e}_r = \mathbf{E}_R, \mathbf{e}_\th = \mathbf{E}_\Theta \right\}$. Transforming the governing equation \eqref{924}, we obtain \cite[see Lemma 2.1 in][]{Norris2008}
\begin{align}\label{927}
  \tilde\nabla^2 \tilde \nabla^2 \eta - {\rho_0 h \omega^2} \eta = 0,
\end{align}
where
\begin{align}\label{928}
  \tilde\nabla^2 =   \lp 1-\f{a^2}{b^2} \rp\lb \f{1}{r} \f{\p}{\p r}\lp \f{r^2-a^2}{r} \f{\p}{\p r}\rp + \f{1}{r^2-a^2} \f{\p^2}{\p \th^2} \rb
\end{align}

Defining the following parameters
\begin{subequations} \label{929}
\begin{align}
  & D'_r = \alpha^2 \mathcal{A}^2(r) D_0,\\
  & D'_\th = \alpha^2 (1/\mathcal{A}(r))^2\ D_0,  \\
  & D'_{r\th} = \alpha^2 D_0, \\
  & \nu'_\th = \f{1}{\alpha^2 \mathcal{A}^2(r)}\lb \mathcal{B}(r) - 4 \log\mathcal{A}(r) \rb,
\end{align}
\end{subequations}
where $\alpha = 1-a^2/b^2$ and $\mathcal{A}(r) = 1-a^2/r^2$ and $\mathcal{B}(r) = 3({r}/{a})\log\lp  \f{r-a}{ r+a}\rp - 2{a^2}/({r^2-a^2})$. We can further expand and simplify equation \eqref{927} as
\begin{align}\label{930}
  & D'_r \f{\p^4 \eta}{\p r^4} + \f{D'_\th}{r^4} \f{\p^4 \eta}{\p \th ^4} + \f{2 D_{r\th}}{r^2}\f{\p^4 \eta}{\p r^2 \p \th^2}+\lp \f{2D'_r}{r} + 2\f{\p D'_r}{\p r} \rp \f{\p^3 \eta}{\p r^3}\nonumber \\
  \quad & + \lp \f{2}{r^2}\f{\p D'_{r\th}}{\p r}-\f{2 D'_{r\th}}{r^3} \rp \f{\p ^3 \eta}{\p r \p \th^2}  + \lb \f{1}{r^2}\f{\p}{\p r}\lp r^2 \f{\p D'_r}{\p r}\rp + \f{1}{r} \f{\p (D_r \nu'_\theta)}{\p r}-\f{D'_\theta}{r^2} +\mathcal{C}(r) \rb \f{\p^2 \eta }{\p r^2} \nonumber \\ 
  \quad & +\lp \f{2D'_{r\theta}}{R^4}  -\f{2}{r^3} \f{\p D'_{r\theta}}{\p r}+ \f{2D'_\theta}{r^4} -\f{1}{r^3} \f{\p D'_\theta}{\p r} +\f{1}{r^2}\f{\p^2 (D'_r \nu'_\theta)}{\p r^2} \rp \f{\p^2\eta}{\p \theta^2}  \nonumber \\
\quad & + \lp \f{D'_\theta}{r^3} -\f{1}{r^2}\f{\p D'_\theta}{\p r} +\f{1}{r} \f{\p^2 (D'_r \nu'_\theta)}{\p r^2}+ \mathcal{D}(r) \rp \f{\p \eta}{\p r} - \rho_0 h \omega^2 \eta =0.
\end{align}
where
\begin{subequations}\label{eq9111}
  \begin{align}
    & \mathcal{C}(r) = \f{1}{2a^2} \lp \f{a}{r}\rp^8 \lb \f{6 - 10(r/a)^2-2(r/a)^4 + {3}(r/a)^6}{ 1 -(a/r)^2} +\f{3}{2}\lp\f{r}{a}\rp^7 \log \lp\f{r-a}{r+a}\rp  \rb D_0,\\
    & \mathcal{D}(r) = \f{3}{a^3}\lp \f{a}{r}\rp^{11}\f{5 -12 (r/a)^2 + 8(r/a)^4}{(1-a^2/r^2)^2}D_0.
  \end{align}
\end{subequations}

Now, we observe that the transformed equation i.e. equation \eqref{930}, matches with the inhomogenious and orhotrpic plate's equation i.e. equation \eqref{923}, with the rigidities and the Poisson ratio as defined in \eqref{929}. The only difference is in the second order term $\p^2\eta/\p r^2$ and the first order term $\p \eta/\p r$ with the coefficients defined in \eqref{eq9111}. Note that these remaining terms goes to zero as long as $r\gg a$, i.e. the panetration depth of the wave into the cloak is small. 

These extra terms in equation \eqref{930} can also be interpreted as an additional pre-stress force $\mathbf{N}$ and body force $\mathbf{S}$ as 
\begin{align}
  N_{rr} = \mathcal{C}(r) , \qquad N_{\th\th} = N_{r\theta} =0, \\
  S_r = -\mathcal{D}(r), \qquad S_\th =0.
\end{align}
Note that the pre-stress force $\mathbf{N}$ and the body force $\mathbf{S}$ satisfy the following constraint as  
\begin{align}
  \n. \mathbf{N} + \mathbf{S} = 0.
\end{align}


\section*{Numerical Solution}

Our cloak is composed of concentric layers of homogeneous materials with a clamped boundary condition at the inner most layer. We have planer incident waves and we aim to find the response of the cloak to the incoming waves. 

For each layer, we expand the solution in that layer using spectral methods as 
\begin{align}
     \eta^{(i)} (r,\th) =  \text{Re} \left\{e^{i\omega t}  \sum_{n=-\infty}^{\infty} \left[ A^{(i)}_n J_n (k_i {r}) + B^{(i)}_n I_n(k_i {r}) + C^{(i)}_n Y_n(k_i {r}) + E^{(i)}_nK_n (k_i r)\right]  e^{ {in\th} } \right\}
\end{align}
where $k_i^4 = \rho h \omega^2/D_i $ with $D_i$ being the flexural rigidity of the layer. Here, $J_n(.), Y_n(.) $ and $I_n(.), K_n(.) $ are respectively Bessel and modified Bessel functions of the first and second kind and $A^{(i)}_n, B^{(i)}_n, C^{(i)}_n, E^{(i)}_n$ are constants that are later found satisfying the boundary conditions.  

Note that outside of the cloak, since the solution should remain finite and satisfy the radiation condition, the solution can be written as 
\begin{align}
     \eta^{\text{out}} (r,\th) =  \text{Re} \left\{e^{i\omega t}  \sum_{n=-\infty}^{\infty} \left[ F_n H^{(1)}_n (k_0 {r}) + G_n K_n(k_0 {r}) + a_0 i^n J_n(k_0 r) \right]  e^{ {in\th} } \right\}
\end{align}
where $H_n^{(1)}(.)$ is the Hankel function of the first kind, $k_i^4 = \rho h \omega^2/D_i $ and $D_i$ being the flexural rigidity outside of the cloak. Note that $a_0 i^n J_n(k_0r)$ represents the planner incident wave. Boundary conditions at the boundary of each layer is continuity of $\eta$, its radial derivative $\eta_r$ and also continuity of momentum and shear force as
\begin{align}
M_r &= -D \left[ \f{\p^2 \eta}{\p r^2} + \nu \lp \f{1}{r} \f{\p \eta}{\p r} + \f{1}{r^2} \f{\p^2 \eta}{\p \th^2} \rp\right] \label{4417} \\
V &= Q_r + \f{1}{r}\f{\p M_{r\th}}{\p \th} \label{4418} 
\end{align}
where
\begin{align}
Q_r &= -D \f{\p}{\p r} \n^2 \eta \label{9310}\\
M_{r\th} &= -D (1-\nu) \f{\p}{\p r}\lp \f{1}{r} \f{\p \eta}{\p \th} \rp \label{9311}
\end{align}

Using the above boundary conditions at each layer and also the clapmed boundary condition $\eta = \eta_r = 0$ at the boundary of the inner most layer, we can solve for the unknowns. 

\section{Discussion on transformation}

In this section we show that the offset $\ep$ introduced earlier, is equivalent to transforming the region $\varepsilon \leq R \leq b$ to the cloaking region $a \leq r\leq b$, where $\varepsilon$ is a small nonzero number. A transformations with a constant Jacobian, mapping the area $\varepsilon \leq R \leq b$ to the cloaking area $a\leq r \leq b$ can be written as 
\begin{eqnarray}  \label{995}
  \mathcal{F}: \left\{
    \begin{array}{ll}
      r=f(R) = \sqrt{c_1 R^2 + c_2}, &\quad  \varepsilon \leq R \leq b, \\
      \th = \Theta,
    \end{array}
\right.
\end{eqnarray}
where 
\begin{align}\label{996}
c_1 = \f{b^2 - a^2}{b^2 - \varepsilon^2}, \qquad  c_2= a^2 - \varepsilon^2 \f{b^2 -a^2 }{b^2 - \varepsilon^2}.
\end{align}
Note that when $\varepsilon =0$, transformation \eqref{995} reduces to the transformation $\mathcal{F}$ in equation \eqref{925}. Picking a value of $\varepsilon/a= 0.175$, we observe that the profile of rigidity becomes the exact same as figure 1.

\bibliographystyle{apalike}
\bibliographystyle{agufull08}

\end{document}